\documentclass[aps,pre,reprint,groupedaddress,noofootinbib,showpacs,floatfix]{revtex4-1}

\usepackage{graphicx}
\usepackage{latexsym}

\usepackage[usenames]{color}

\begin{document}

\title{Non-Markovian dynamics of reaction coordinate in polymer folding}

\author{T. Sakaue}
\affiliation{Department of Physics, Kyushu University, Fukuoka 819-0395, Japan}
\affiliation{JST, PRESTO, 4-1-8 Honcho Kawaguchi, Saitama 332-0012, Japan}
\author{J.-C. Walter}
\affiliation{Laboratoire Charles Coulomb, UMR5221 CNRS-UM,
Universit\'e de Montpellier, Place Eug\`ene Bataillon, 34095 Montpellier
Cedex 5, France}

\author{E. Carlon}
\affiliation{Institute for Theoretical Physics, KU Leuven, Celestijnenlaan
200D, B-3001 Leuven, Belgium}

\author{C. Vanderzande}
\affiliation{Faculty of Sciences, Hasselt University, Agoralaan 1, B-3590 Diepenbeek, Belgium}
\affiliation{Institute for Theoretical Physics, KU Leuven, Celestijnenlaan
200D, B-3001 Leuven, Belgium}

\def\degC{\kern-.2em\r{}\kern-.3em C}
\def\SimIneA{\hspace{0.3em}\raisebox{0.4ex}{$<$}\hspace{-0.75em}\raisebox{-.7ex}{$\sim$}\hspace{0.3em}} 
\def\SimIneB{\hspace{0.3em}\raisebox{0.4ex}{$>$}\hspace{-0.75em}\raisebox{-.7ex}{$\sim$}\hspace{0.3em}} 

\date{\today}

\begin{abstract}
We develop a theoretical description of the critical zipping dynamics
of a self-folding polymer.  We use tension propagation theory and the
formalism of the generalized Langevin equation applied to a polymer that
contains two complementary parts which can bind to each other. At the
critical temperature, the (un)zipping is unbiased and the two strands
open and close as a zipper. The number of broken base pairs $n(t)$
displays a subdiffusive motion characterized by a variance growing as
$\langle \Delta n^2(t) \rangle \sim t^\alpha$ with $\alpha < 1$ at long
times. Our theory provides an estimate of both the asymptotic anomalous
exponent $\alpha$ and of the subleading correction term, which are both
in excellent agreement with numerical simulations.  The results indicate
that the tension propagation theory captures the relevant features of
the dynamics and shed some new insights on related polymer problems
characterized by anomalous dynamical behavior.  
\end{abstract}

\maketitle
\section{Introduction}

Conformational dynamics of biopolymers, such as DNA, RNA and proteins,
is a complex process involving a large number of degrees of freedom. Like
any other many-body problem, the concept of the {\it reaction coordinate}
(RC) is often invoked in its coarse grained description. One may be
tempted to assume Markovian dynamics for the RC such that the problem
is amenable to standard stochastic analysis~\cite{vank95}. 
However, the validity of such
a simple 
approach
requires that the RC is the slowest variable 
and that its characteristic
time scale is
well separated from 
all other time scales in the problem. 
This condition is not easily met 
in many situations, giving rise to non-Markovian effects 
and anomalous dynamics.

Anomalous diffusion is an ubiquitous phenomenon observed in a
large number of experimental systems or in computer simulations
\cite{mand68,bouc90,ambl96,krug97,metz00,panj09,amit10,liza10,akim11,metz14}.
Characteristic of these systems is a mean squared displacement (MSD)
of particle positions (or more generally of some RC) which scales
asymptotically in time as $\langle \Delta {\vec x}^2 (t) \rangle \sim
t^{\alpha}$ with $\alpha \neq 1$, i.e. deviating from the Brownian
motion predictions. The evidence of anomalous dynamics is mostly,
both in experiments and simulations, of observational/empirical nature.
Due to the complexity of the systems studied it is hard to predict the
value of $\alpha$ from theoretical inputs.

In this paper, we investigate the anomalous diffusion of the RC in
a simple system with folding dynamics: the (un)zipping in hairpin
forming polymers~\cite{mang16}. In this process the polymer contains two
complementary parts which can bind to each other and fluctuates between
an open (unzipped) and a closed (zipped) conformation. We focus here on
the dynamics at the transition temperature where zipped and unzipped
state have the same equilibrium free energy. The natural RC for the
system is the number of broken base pairs $n(t)$. The time series of
$n(t)$ exhibits back and forth fluctuations reminiscent to Brownian
motion. Simulations of the mean-square displacement (MSD) reveals the
motion is sub-diffusive $\langle \Delta n^2(t) \rangle \sim t^{\alpha}$
with $\alpha < 1$~\cite{walt12}.

Here, we clarify the non-Markovian nature of this process using an
analysis of the collective dynamics of the polymer, based on the tension
propagation along the polymer backbone.  A perturbation propagates
along the backbone due to the tension transmitted along the chain,
generating long range temporal correlations. The theory enables us to
provide an analytical estimate of $\alpha$ including the sub-leading term.
Our predictions are in very good agreement with the results of computer
simulations, which demonstrates the validity of our approach and sheds new
insight on related polymer problems characterized by anomalous diffusion.

The theory is based on the Generalized Langevin Equation (GLE) formalism,
which is briefly reviewed in Sec.~\ref{sec:GLE}. The key point is the
calculation of the memory kernel entering in the GLE and characterizing
the non-Markovian aspects of the dynamics. This calculation is done
in Sec.~\ref{sec:kernel} and allows to estimate both the leading
exponent $\alpha$ and the subleading term. In Sec.~\ref{sec:numerics}
we show that the analytical predictions are in excellent agreement
with numerical simulations of the (un)zipping process. Finally, in
Sec.~\ref{sec:conclusion}, we present our conclusions and we point out
the relation of our results to the problems of tagged monomer motion
and polymer translocation.

\section{Generalized Langevin Equation}
\label{sec:GLE}

Consider a step displacement applied to an appropriate RC ${\vec z}(t)$.
Let us
monitor the subsequent
average force ${\vec f}(t)$ to keep the given displacement. This protocol
can be analyzed by the force balance equation
\begin{eqnarray}
\int_{t_0}^t d\tau \ \gamma(t-\tau) {\vec v}(\tau) = {\vec f}(t)
\label{force_balance_memory}
\end{eqnarray} 
where ${\vec v}(t)=d{\vec z}(t)/dt$ and $\gamma(t)$ is the
memory kernel (in Markovian systems $\gamma(t)\sim \delta(t)$).
In Eq.~(\ref{force_balance_memory}) we may set the lower bound of the time
integral as $t_0 \rightarrow -\infty$ by assuming the system is already
in the equilibrium state before the operation is made.  In the case of a
step displacement ${\vec u}$ imposed at $t=0$, i.e., ${\vec z}(t+0) =
{\vec z}(t-0)+{\vec u}$, we have ${\vec v}(t) = {\vec u} \delta(t)$,
the above equation is reduced to
\begin{eqnarray}
u \gamma(t) = f(t)
\label{stress_relaxation}
\end{eqnarray}
where we have switched to a scalar notation by noting ${\vec u}
\parallel {\vec f}$ in isotropic system.  

To connect the average stress relaxation with the anomalous
fluctuating dynamics, we need to look at each realization of
the stochastic processes
by adding
the thermal noise term ${\vec \xi}(t)$ to the right-hand side
of Eq.~(\ref{force_balance_memory}). The noise has zero mean $\langle
\xi_{\text i}(t) \rangle =0$, and it is related to the memory kernel via
the fluctuation-dissipation theorem (FDT) $\langle \xi_{\text i}(t)
\xi_{\text i}(\tau) \rangle = k_BT \gamma(|t-\tau|) \delta_{\text ij}$. The
equivalent expression of the Generalized Langevin Equation (GLE) is
\begin{eqnarray}
{\vec v}(t) = \int_{t_0}^{t} d\tau \ \mu(t-\tau) {\vec f}(\tau) +  
{\vec \eta}(t)
\label{GLE_2}
\end{eqnarray}
where $\mu(t)$ is the mobility kernel with the FDT $\langle \eta_{\text
i}(t) \eta_{\text j}(\tau) \rangle = k_BT \mu(|t-\tau|) \delta_{\text ij}$ \cite{sait15,panj10,vand17,saka16}. 
In the next section a power-law decaying memory function $\gamma(t)
\sim t^{-\alpha}$ in the case of polymer pulling is derived from polymer tension propagation
arguments. From this one derives $\mu(t) \sim -t^{\alpha-2}$ (for details
see Appendix~\ref{sec:app1}).  In the unbiased case ${\vec f}(t)=0$,
the MSD can be derived after integration of the velocity correlation
function $\langle {\vec v}(t) \cdot {\vec v}(s) \rangle = \langle {\vec
\eta}(t) \cdot {\vec \eta}(s) \rangle$ twice with respect to time,
yielding $\langle \Delta {\vec z}(t)^2 \rangle \sim t ^{ \alpha}$, i.e.,
the stress relaxation exponent characterizing the decay of the
memory kernel $\gamma(t)$ is equal to the MSD exponent.

%

\begin{figure*}[t]
\includegraphics[width=0.95\textwidth]{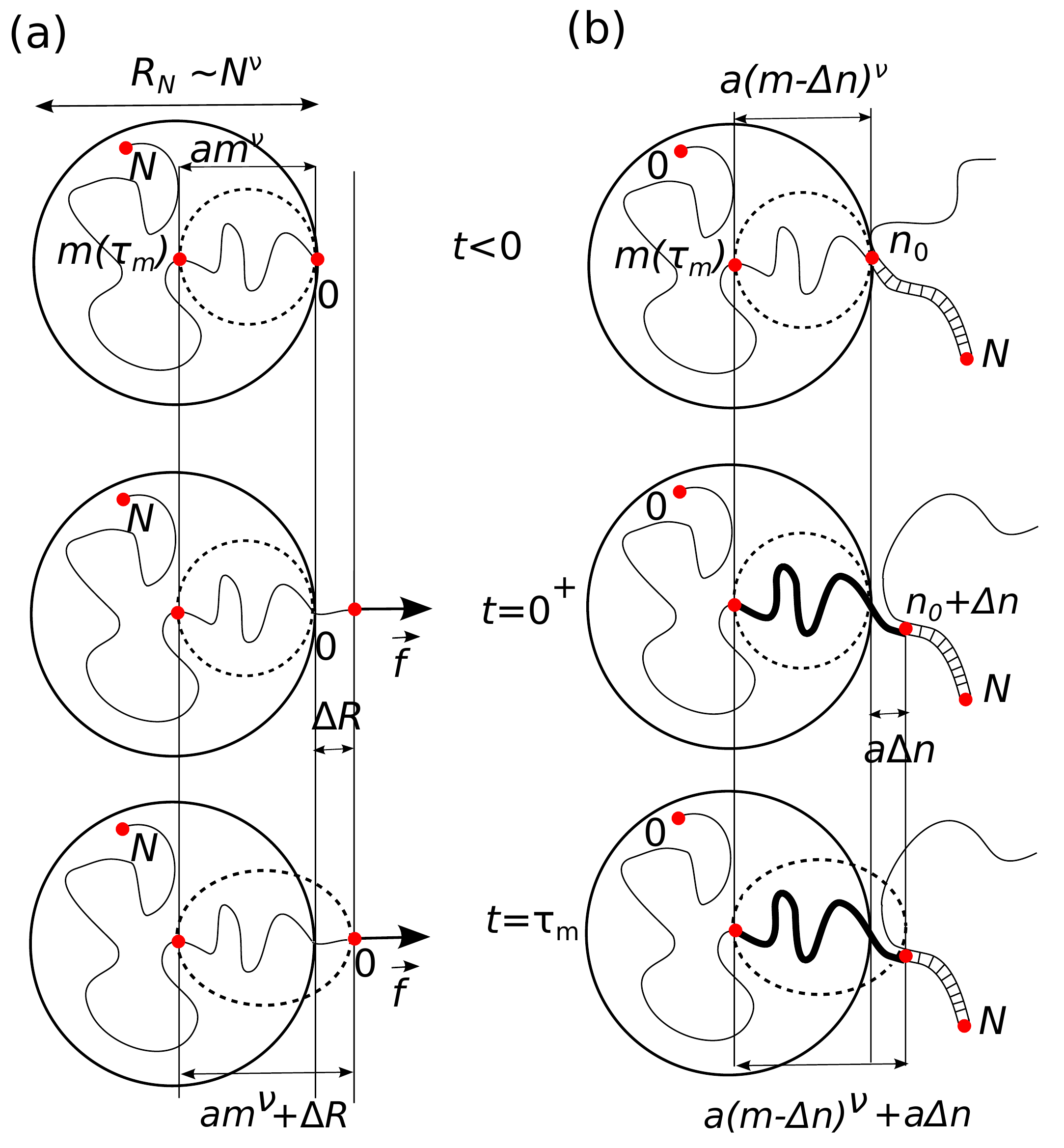}
\caption{
Illustration of the differences in tension propagation
between a polymer pulled by one end (a) and a zipping polymer (b).  In
both cases to calculate the memory kernel one starts from an equilibrated
conformation and introduces a perturbation.  (a) Pulling occurs at time
$t=0$. At time $t=\tau_m$ the tension front has reached the monomer
$m$. Here the deformation $\Delta R$ is fixed by the initial pulling.
(b) In zipping peeling of $\Delta n$ monomers occurs at a time $t=0$.
At a time $t=\tau_m$ the perturbed part of the chain involves $m(t)$
monomers counting from the new fork point (thick line in the figure). The
deformation $\Delta R_m$ of the polymer is equal to the difference
between the actual size of this perturbed part and its equilibrium size
(Eq.~\ref{d_R}).
}
\label{Fig01}
\end{figure*}

\section{Memory kernel for zipping dynamics}
\label{sec:kernel}

Before dealing with the more complex case of zipping polymers, it is
useful to recall some known results \cite{saka12,sait15} for a simpler
case of polymer pulling (see Fig.~\ref{Fig01}(a)). Let us suppose that
one end of an equilibrated polymer is displaced by $\Delta x$ at $t=0$ and
that the position of that monomer is kept fixed. This operation produces
a stretching of the end part of the chain. Through tension propagation
the polymer relaxes to a new equilibrium state shifted with respect to
the original position. The longest relaxation time is $\tau_N \simeq
\tau_0 N^{z\nu}$, where $\tau_0$ is a monomer time scale, $\nu\simeq 0.588$
is the Flory exponent and the $z=2 + 1/\nu$ is the dynamical exponent
(we consider here the free draining case, if hydrodynamic interactions
are taken into account $z=3$). At a time $t < \tau_N$ only $m(t)$
monomer close to the displaced end are stretched, while the remaining
$N-m(t)$ at the opposite end do not yet feel the displacement operation.
The longest relaxation time for a fragment containing $m$ monomers is
$\tau_m \simeq \tau_0 m^{\nu z}$, from which one finds
\begin{eqnarray}
m(t) &\simeq& \left(\frac{t}{\tau_0} \right)^{\frac{1}{\nu z}}
\label{tension_propagation}
\end{eqnarray}
which gives how $m$ grows in time.
To keep the end monomer at a fixed position one needs to apply
a force $f(t)$ which can be estimated using polymer entropic 
elasticity. An equilibrated polymer stretched by $\Delta x$ exerts
a force at its two ends which is equal to:
\begin{eqnarray}
f &\simeq& \frac{k_BT}{\langle R^2 \rangle} \Delta x 
\label{forceR}
\end{eqnarray}
where $\langle R^2 \rangle$ indicates the average of the squared
end-to-end distance. Applying the previous relation to the stretched
$m(t)$ monomers, for which $\langle R^2 \rangle \simeq a^2 m^{2\nu}$
and using Eq.~(\ref{tension_propagation}) we obtain
\begin{eqnarray}
\gamma (t) = \frac{f(t)}{\Delta x} &\simeq& 
\frac{k_BT}{a^2} \left( \frac{t}{\tau_0} \right)^{-2/z} 
\label{gamma_t}
\end{eqnarray}
where we used Eq.~(\ref{stress_relaxation}) for a step displacement
equal to $\Delta x$.  Equation~(\ref{gamma_t}) gives the memory kernel
associated to the step displacement of a polymer end. According to the
discussion of the previous section the decay exponent of $\gamma(t)$
is equal to the MSD exponent. Hence we obtain $\alpha = 2/z$.  For an
ideal Rouse chain for which $z=4$ ($\nu=1/2$), one obtains a tagged
monomer diffusion with MSD scaling as $\Delta \vec{x}^{\, 2}(t) \sim
t^{1/2}$, which is in agreement with the exact solution from Rouse
dynamics~\cite{doi88}.  More generally the tension propagation dynamics
leads to a subdiffusive behavior with $\alpha = 2\nu/(1+2\nu) < 1$,
which turns into ordinary diffusion at times $t > \tau_R$.

We turn now to the case of zipping dynamics. Let us assume that
the polymer is in equilibrium with $n_0$ bonds from the tail being in
unzipped state, while the remaining $N-n_0$ bonds are zipped, i.e., the
monomer's label at the fork point is $n(t) = n_0$ ($t < 0$).  Consider now
an instantaneous break of $\Delta n (= {\mathcal O}(1))$ zipped pairs
at the fork point creating $\Delta n $ additional unzipped monomer
pairs. This operation produces (i) the change of the reaction coordinate
$n(t) = n_0 \rightarrow n_0 + \Delta n$ and (ii) the displacement of
the position of the fork in real space ${\vec r}(n_0, 0) \rightarrow
{\vec r}(n_0+\Delta n, 0)$, where ${\vec r}(n,t)$ is the position of the
monomer $n$ at time $t$ (Fig.~\ref{Fig01}(b)).  As in the pulling problem the
entire chain cannot respond to the break of $\Delta n$ bonds all at once.
At time $t$ smaller than the longest relaxation time of the polymer
only a finite section, i.e., $m(t)$ bonds given close to the fork point
respond to the perturbation. 

The deformation of such a responding part of the chain can be evaluated as
(Fig.~\ref{Fig01}(b))
\begin{eqnarray}
\Delta R_m(t) &\simeq&  a \Delta n + a \{m(t)-\Delta n\}^{\nu}- a m(t)^{\nu} 
\nonumber \\
&\simeq& a[\Delta n - \nu \Delta n m(t)^{\nu-1}] 
\label{d_R}
\end{eqnarray}
where we have taken $m(t) \gg \Delta n$ and expanded to lowest order in
$\Delta n$.  The previous equation can be understood as
follows. There are $m(t)$ monomers in the part of the unzipped arm which
is under tension (thick line in Fig.~\ref{Fig01}(b)). The equilibrium
radius of this part would be $a m(t)^{\nu}$. However at time $t$
the actual size is $a \Delta n + a \{m(t)-\Delta n\}^{\nu}$
because the average position of the monomer at the tension front is
not yet affected by the peeling at this time scale. The total size is
the sum of the unperturbed size of $m(t)-\Delta n$ monomers and of the
peeled part which is $a \Delta n$.
The deformation $\Delta R_m(t)$ is then obtained by
subtracting the actual radius of the $m(t)$ monomers and the
equilibrium value, which leads  to Eq.~(\ref{d_R}).

The growth of $m(t)$ in time is governed
by the tension propagation dynamics of Eq.~(\ref{tension_propagation}).
The force necessary to hold the fork point to the new position $n_0 +
\Delta n$ can be estimated again from entropic elasticity (Eq.~(\ref{forceR})) as
\begin{eqnarray}
f(t) &\simeq& \frac{k_BT}{\langle R^2(t) \rangle}\Delta R_m(t) 
\nonumber \\
&\simeq& \frac{ k_BT}{a} \Delta n 
\left[ \left(\frac{t}{\tau_0}\right)^{-2/z} - 
\nu \left( \frac{t}{\tau_0}\right)^{-\frac{1+\nu}{\nu z}}\right]
\label{ft_zipp}
\end{eqnarray}
Dividing by $a\Delta n$ we obtain the memory kernel with a leading $t$
behavior as in Eq.~(\ref{gamma_t}), but now the analysis unveils the
presence of a sub-leading term. The calculation of the MSD which follows
from Eq.~(\ref{ft_zipp}) is given in the Appendix~\ref{sec:app1},
where the full calculation of $\gamma(t)$ is presented including the
subleading term.  The final result for the RC dynamics is
\begin{equation}
\langle \Delta n^2(t) \rangle \sim t^{2/(\nu z)}
\left( 1 + C t^{-(1-\nu)/(\nu z)}\right)
\label{alpha_alpha1}
\end{equation}
with $C$ a positive constant.

\begin{figure}[t]
{
\includegraphics[width=0.4\textwidth]{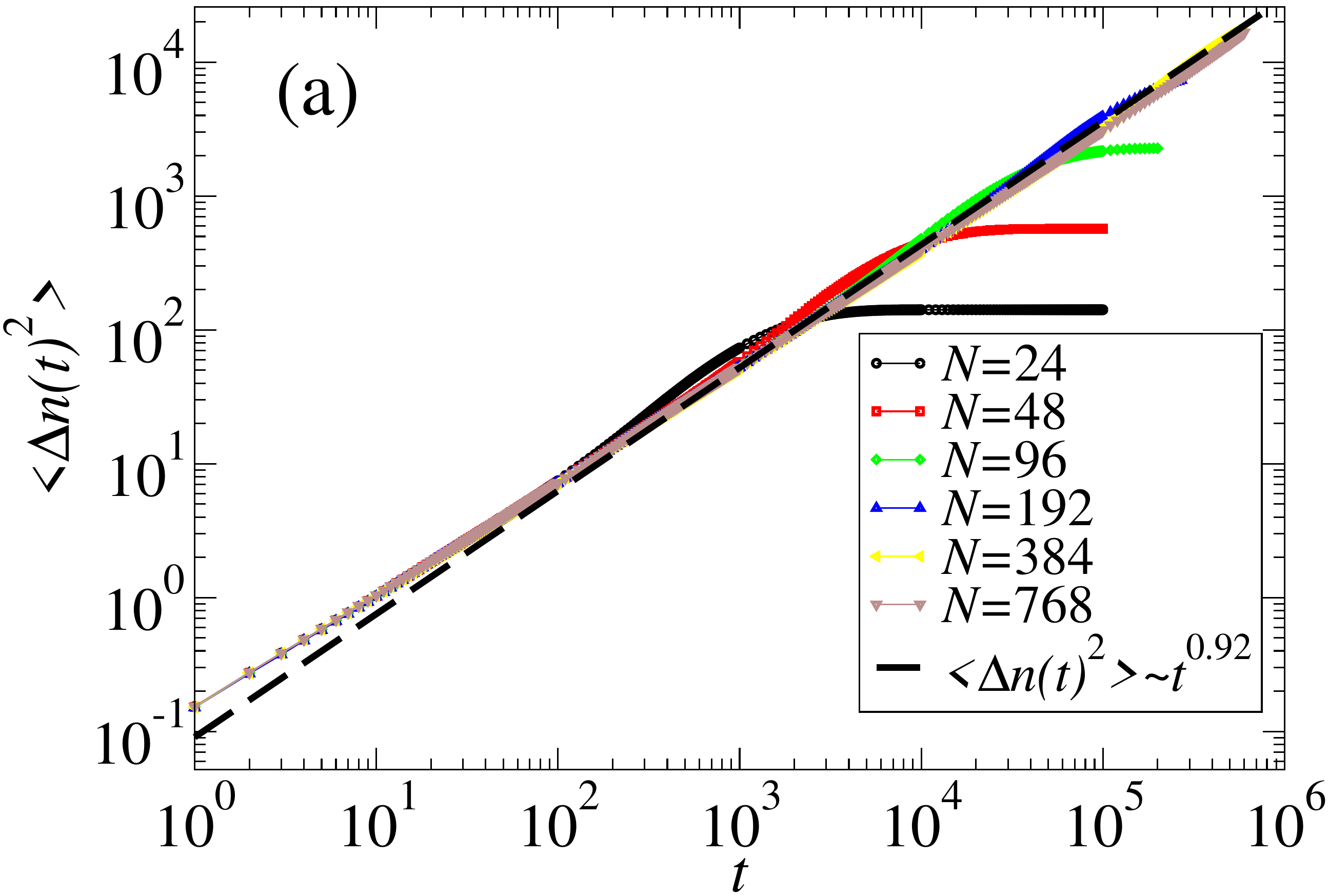}
\includegraphics[width=0.4\textwidth]{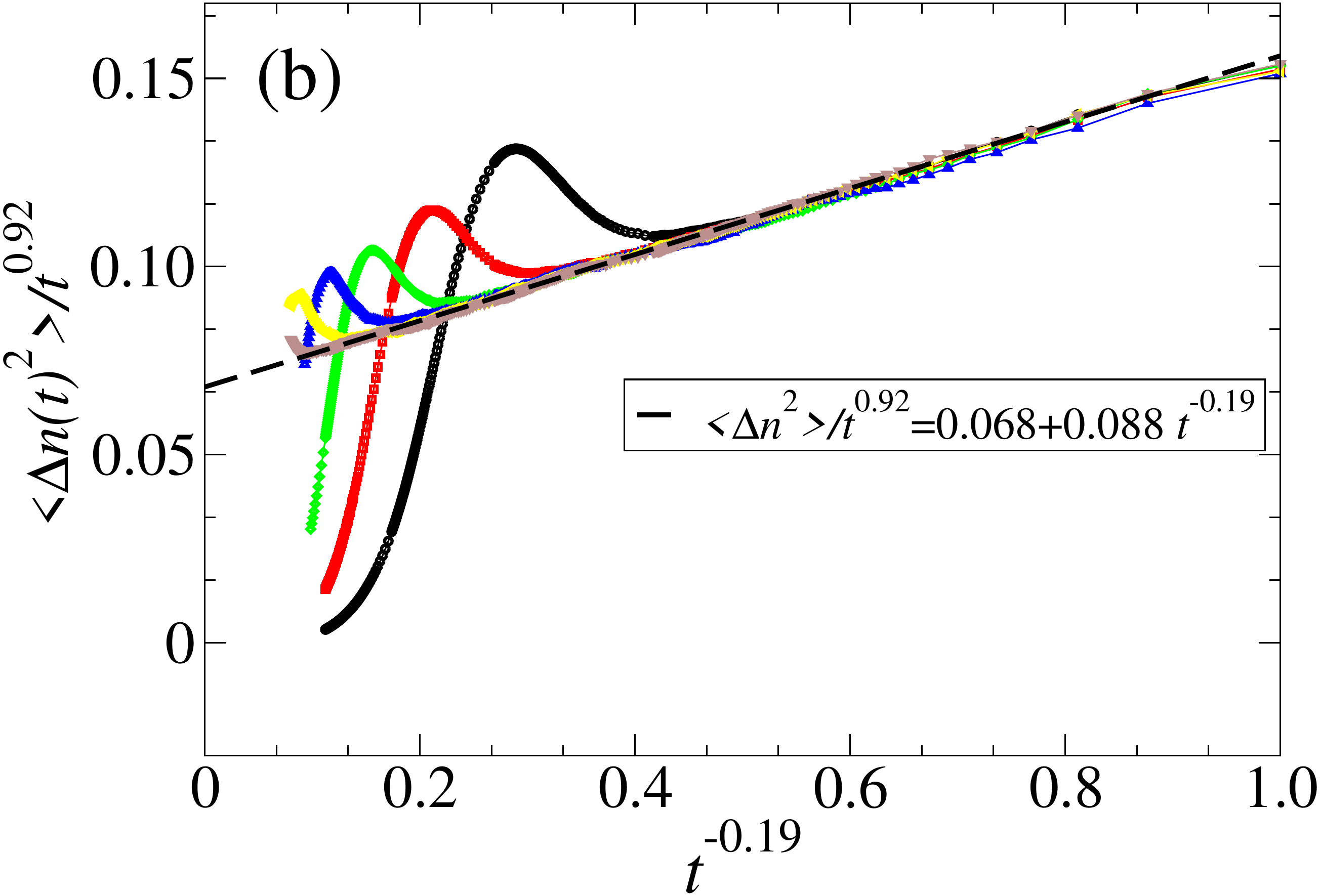}
}
\caption{(a) The mean-square displacement $<\Delta n(t)^2>$ of the
reaction coordinate (fork location along the chain) is plotted versus
time for different sizes.  The symbols are obtained for simulations
with different sizes (from~\cite{walt12}). The dashed line is the
theoretical prediction with an exponent 0.92.
 The deviation of the MSD from the leading term at short time is explained by the subleading term in Eq.(9).
 (b) Correction to
scaling $\Delta n(t)^2/t^{0.92}$ plotted versus $t^{-0.19}$.  According
to Eq.(\ref{alpha_alpha1}), the resulting curve should fit linearly. 
Remarkably, even the numerical value of the slope of the corrections 
are in good agreement with theory. 
For both the leading and first order correction to scaling, the numerics
are in very good agreement with theoretical predictions.}
\label{FigMSD}
\end{figure}

\section{Numerical Results}
\label{sec:numerics}

The model used in the simulations is discussed in details in
Ref.~\cite{walt12} and was also employed in previous studies of
renaturation dynamics~\cite{ferr10}. We consider two strands with $N$
monomers which are joined to a common monomer, labeled with $i=N$,
while we use an index $i=1 \ldots N$ to label the monomers on the two
strands.  Only monomers with the same index $i$ on the two strands can
bind with binding energy $\varepsilon$. The dynamics consists of lattice
corner-flips or end-flips local moves which are randomly generated by
a Monte Carlo algorithm.  This algorithm was shown to reproduce the
Rouse model dynamics in previous studies~\cite{ferr11} and represents an
interesting and efficient alternative to the more commonly used Langevin
dynamics for polymers in the continuum.

A Monte Carlo move not respecting mutual or self-avoidance between the two
strands is rejected. A move binding two monomers on the opposite strands
is always accepted, while the opposite move of unbinding is accepted
with a probability $\exp(-\beta\varepsilon)<1$, where $\beta=1/k_BT$ is
the inverse temperature. The algorithm hence satisfies detailed balance.
The temperature is tuned to the critical value $\beta=\beta_c$, which
is very accurately known as it relies on previous high precision data
about polymers on an fcc lattices~\cite{ishi89}. In addition in the model
bubbles are not allowed to form so the dynamics is strictly sequential
as in a zipper.

\begin{figure}[t]
\includegraphics[width=0.4\textwidth]{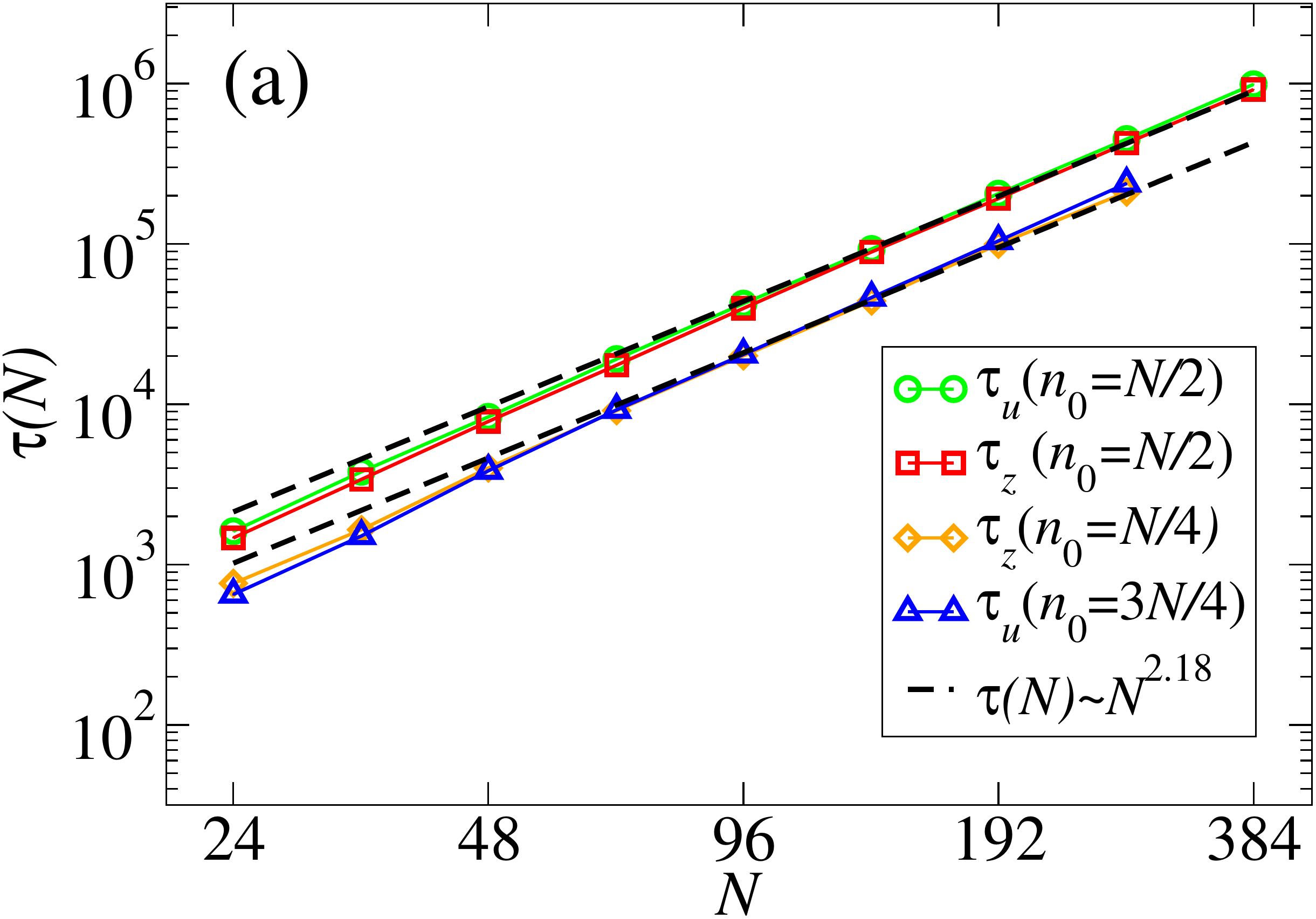}
\includegraphics[width=0.4\textwidth]{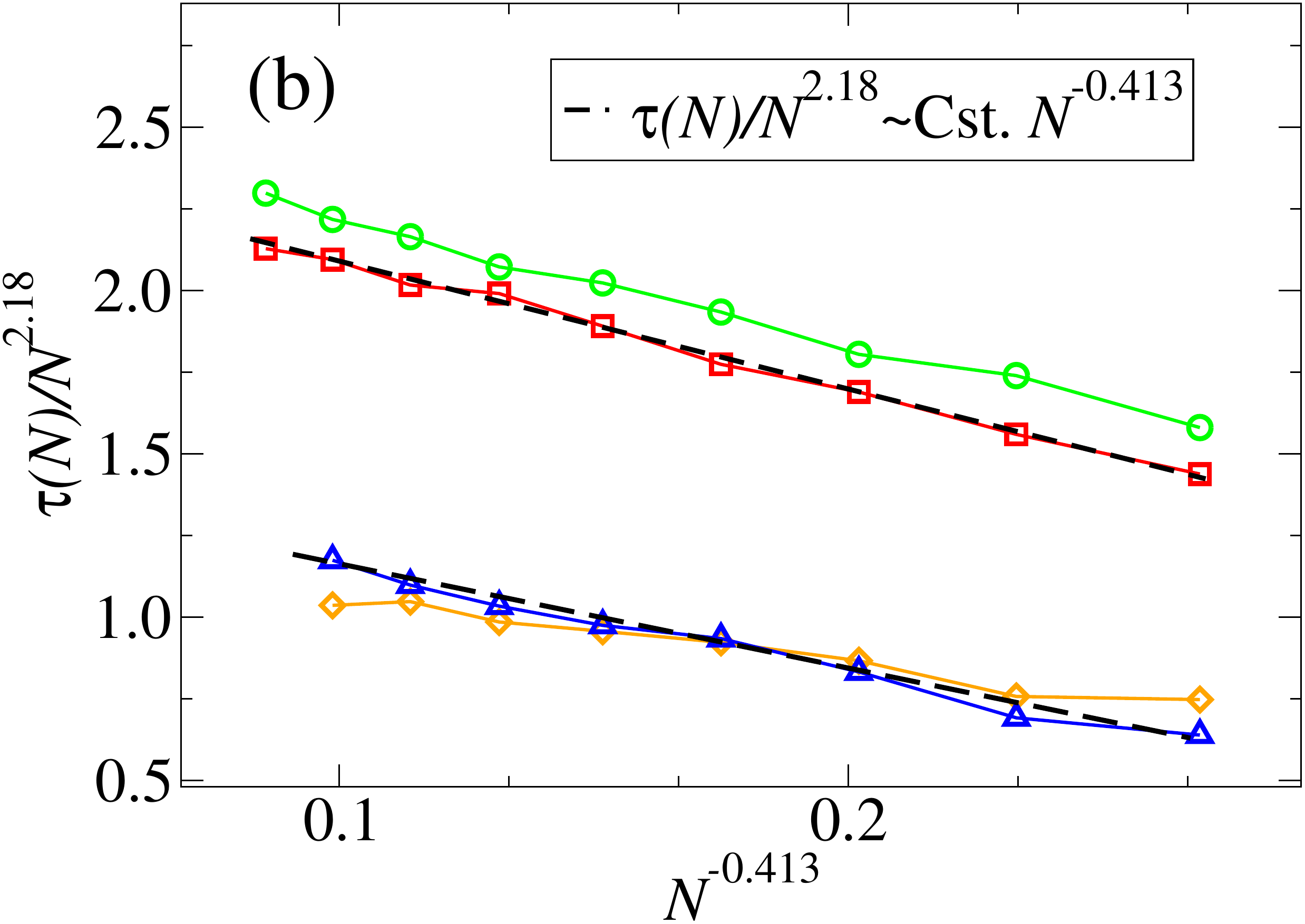}
\caption{(a) The zipping and unzipping time $\tau_z$ and $\tau_u$,
 respectively, are plotted versus the polymer length
for different initial conditions ($n_0=N/4,$ $N/2$ and $3N/4$). Symbols are
results from simulations, and Eq.(\ref{scaling_tau}).  The analytical
estimate of the exponent $2.18$ is in good agreement with the numerics
for large sizes. (b) $\tau/N^{2.18}$ is plotted versus the
first order correction to scaling $t^{-0.413}$.  The resulting curve is
expected to be a straight line with a negative slope.  Both leading term
and first order correction are in good agreement with the numerics, 
independent from the initial value of $n_0$.}
\label{FigTau}
\end{figure}

A simulation run is initialized by setting the fork point to
$n(t=0)=N/2$, so that monomers $0 \leq i \leq N/2$ are unbound and 
$i > N/2$ are bound. The initial configuration is equilibrated by
sufficiently long Monte Carlo runs
 while keeping the
fork point fixed (see Appendix \ref{sec:sim_num}).
 After equilibration, the constraint is released and
the actual simulation is started. The fork point performs a stochastic
back and forth motion along the polymer backbone until one
 of the two ends is reached and the simulation is stopped.
  We monitor in particular
the MSD $\langle \Delta n^2(t) \rangle$ and the average duration time
of the process $\tau$.

The analysis of Ref.~\cite{walt12} showed that the dynamics is
well-described by a fractional brownian motion (fBm) characterized by
a Hurst exponent $H=0.44(1)$ (recall that in fBM the Hurst exponent
is linked to the MSD exponent by the relation $\alpha=2H=0.88(2)$
and that the fBm is described by a GLE).  The analytical
prediction of Eq.~(\ref{alpha_alpha1}) is $\alpha=2/(\nu
z)=0.92$, which is somewhat higher that the numerical value of
Ref.~\cite{walt12}. Figure~\ref{FigMSD}(a) shows a plot of the MSD
for lattice polymers of lengths up to $N=768$ and averaged over
$\sim 5\cdot10^5$ realizations.  The dashed line in Fig.~\ref{FigMSD}(a) is
the analytical prediction. The data converge to this prediction for
sufficiently long times, with some deviations close to the saturation
level (obviously the MSD cannot grow beyond the squared half total length
of the strands).  At short times there is a visible deviation from the
analytical prediction.

In order to test the validity of Eq.~(\ref{alpha_alpha1}) we plot
in Fig.~\ref{FigMSD}(b) the quantity $\langle \Delta n^2(t) \rangle
t^{-0.92}$ vs. $t^{-(1-\nu)/(\nu z)}=t^{-0.19}$.
The MSD plotted in these rescaled
unit is expected to show a linear behavior, which is indeed observed
in Fig.~\ref{FigMSD}(b). Also it is important to note that the theory
predicts a positive coefficient $C>0$ in Eq.~(\ref{alpha_alpha1}),
as discussed in Appendix~\ref{sec:app1}, and this is indeed consistent
with the numerics. Moreover, a fit of the correction gives 
$\langle n(t)^2\rangle/t^{0.92}=0.068(1+C\,t^{-0.19})$ 
where the prefactor $C=1.294$ is 
in good agreement with the theoretical
 prediction $1.726$ given by Eq.(\ref{final}). 
Hence we can conclude that the numerical data are in
excellent agreement with the tension propagation theory predictions.



Additional support to the theory is obtained from the analysis of the
average  time $\tau$ to fully zip or unzip as a function of the polymer length $N$,
 see Fig.\ref{FigTau}.
 The system is prepared in different initial conditions.
 For $n_0=N/2$, we sample both the unzipping time $\tau_u$ and the 
zipping time $\tau_z$ depending which end $N$ or 0 is reached first,
 respectively. We also sample the zipping time for $n_0=N/4$,
 and the unzipping time for $n_0=3N/4$. 
This time is expected to be an increasing
function of the strands length $N$. From Eq.(\ref{alpha_alpha1}) one
obtains the asymptotic scaling $\tau \sim N^{\nu z}$. This result can be extended to the next
order correction from the analysis of Eq.~(\ref{alpha_alpha1})
\begin{equation}
\tau \sim N^{\nu z} \left( 1 - D N^{-(1-\nu)} +\ldots\right)
\label{scaling_tau}
\end{equation}
with $D>0$. To confirm it, we first present
a log-log plot of $\tau$ vs. $N$ in Fig.~\ref{FigTau}(a), which shows some
deviations from the asymptotic behavior $\tau \sim N^{\nu z}$.
 We then show a plot of $\tau N^{-\nu z}$ vs. $N^{-(1-\nu)}$ in Fig.~\ref{FigTau}(b).
The rescaled data follow a straight line with a negative slope in very
good agreement with the prediction of Eq.~(\ref{scaling_tau}). 
We do not observe any change in the dynamical scaling
 for different initial conditions as it has been observed in the related system
 of protein search on DNA \cite{lang15}.

\section{Conclusion}
\label{sec:conclusion}

The anomalous dynamics in polymers originates from growth of the cooperatively moving domain, i.e., tension propagation along the chain:
 a perturbation on a given position propagates along the
polymer backbone creating a viscoelastic memory effect for the motion
of individual monomers. A theoretical framework of tension propagation has been mostly
 developed in the past years for the nonequilibrium dynamics of polymers ,i.e. the analysis of driven polymer translocation~\cite{saka07,saka10,ikon12}
 and the polymer stretching process~\cite{saka12,sait15,rowg12,vand14,vand17}.
In near-equilibrium (or unbiased) situations, the essential physics is also given by
 the growing length scale of the cooperative motion as the source of anomalous dynamics,
 for which the scaling form of the tension propagation is determined by an equilibrium
 argument, see Eq.\ref{tension_propagation}.
 In the unbiased translocation dynamics it is a monomer exchange across the pore that generates
 a long range decay of the memory kernel \cite{panj07,panj10,saka16}, while in the tagged monomer motion the same effect
 is due to the spatial displacement by pulling \cite{sait15,panj10,saka13}.

The (un)zipping dynamics analyzed in this paper can be understood as a hybrid of the above two processes. It is the monomer exchange
$\Delta n$ (cf. Fig.~\ref{Fig01}) between zipped and unzipped sections 
which creates a long range temporal memory
leading to a power-law decaying memory kernel as in Eq.~(\ref{ft_zipp}).
Inspecting the elementary process, we see that the first term in the RHS of
Eq.~(\ref{d_R}) reflects the process entailing the spatial displacement ${\vec r}(n_0, 0) \rightarrow
{\vec r}(n_0+\Delta n, 0)$,
while the second term concerns the change in
$\Delta n$ without spatial displacement. The latter is reminiscent to the translocation process entailing
 the monomer exchange across the pore, while the spatial position of the RC is fixed at the pore site.

The present formalism enabled us to extract the anomalous diffusion
characteristics of the RC including the subleading behavior
\begin{equation}
\langle \Delta n^2(t) \rangle \sim 
t^{\alpha} \left( 1 + C t^{-\alpha_1} + \ldots\right)
\end{equation}
with analytical expressions for $\alpha$ and $\alpha_1$ which are found to
match very well the numerical simulation data.  Since the dominant source
of the tension generation comes from the spatial displacement of the RC,
a process equivalent to pulling operation (the first term in the RHS of
Eq. (\ref{d_R})), the asymptotic anomalous diffusion exponent $\alpha
=2/(\nu z)$ is controlled by that of the tagged monomer diffusion,
see Eq.~(\ref{ft_zipp}), while the
subleading exponent $\alpha - \alpha_1 =(1+\nu)/(\nu z)$ coincides
with that expected for the unbiased polymer translocation (see Eq.~(9)
in Ref.~\cite{saka16}).  Note, however, that the translocation problem
is complex because of a series of factors (post-propagation behavior,
interaction with the pore), and simulations, at least in the unbiased
case, are still controversial~\cite{chua01,haan12,paly14,saka16}.

From a broader perspective, we repeat once more the caution on the RC
based coarse grained description. The validity of the assumption leading
to the Markovian dynamics is generally dependent on the time scale at
hand (say, observation), but as we have shown here, there would exist for
the dynamics of long polymers a broad time window, in which collective
dynamics among degrees of freedom with varying time scale manifests. 
Indeed, the Markovian description is valid only on the time scale coarser
 than the longest relaxation time $ \tau_N \simeq \tau_0 N^{\nu z}$ of the molecule.
 Therefore, the slow dynamics is a generic feature in high molecular weight macromolecules,
 and this implies that on the time scale $(t < \tau_N)$ relevant to the conformational dynamics,
 only the partial section of a chain can be equilibirated. Our present theory utilizes
 equilibrium properties of such an equilibriated section, whose
 size evolves in time along with the tension propagation. This allows us to clarify the stress
 relaxation and the anomalous dynamics of RC due to the viscoelastic response.
The resulting non-Markovian dynamics should be of pronounced importance
in the context of biopolymer functions. Although more work is necessary
to fully unveil the consequences, our analytical argument for the MSD
is regarded a first step toward such an ambitious goal.

Away from the critical point, at low temperatures, the hairpin folding
process exhibits out-of-equilibrium characteristics~\cite{fred14} which resembles
scaling behavior observed in DNA hairpin experiments \cite{neup12}.
That case is reminiscent of polymer translocation driven by external bias
~\cite{saka07,saka10,letho09,bhat10,ikon12,saka16}.
Here again, a key physics lies in the tension propagation,
the dynamics of which bears distinctive features not seen
in the unbiased regime discussed in this work.



\begin{acknowledgements}
This work is supported by KAKENHI (No. 16H00804, “Fluctuation and
Structure”) from MEXT, Japan, and JST, PREST0 (JPMJPR16N5).  This work is also part
of the program Labex NUMEV (AAP 2013-2-005, 2015-2-055, 2016-1-024).
\end{acknowledgements}
\appendix*

\section{The correction to scaling behavior}
\label{sec:app1}

%
%

We give here the full derivation of the calculation of the MSD including
the subleading corrections. The calculation consists of two steps.
Firstly we determine the mobility kernel $\mu(t)$ and from it, 
using the FDT, we obtain the MSD.

\subsection{Mobility Kernel}

Taking the Laplace transforms of Eqs.~(\ref{force_balance_memory}) and
(\ref{GLE_2}) one obtains the following relation:
\begin{equation}
\hat{\mu}(s) = \frac{1}{\hat{\gamma}(s)}
\label{eq:hatmu}
\end{equation}
(generalizing the relation between mobility and friction).  In the
previous equation $\hat{\gamma}(s)$ and $\hat{\mu}(s)$ are the Laplace
transforms of $\gamma(t)$ and $\mu(t)$, respectively.
In what follows we calculate the Laplace transform of the memory
kernel $\hat\gamma(s)$ and then obtain $\hat{\mu}(s)$ from Eq.~(\ref{eq:hatmu}).
Finally we use the inverse Laplace transform to obtain
$\mu(t)$. This can be readily done for a pure power law function
$\gamma(t)=t^{-\alpha}$: its Laplace transform is $\hat\gamma(s)=
\Gamma(1-\alpha) s^{\alpha-1}$, where $\Gamma(z)$ is the Euler gamma
function.  Therefore, neglecting the prefactor, $\hat\mu(s)\sim
s^{1-\alpha}$ which leads to $\mu(t) \sim t^{\alpha-2}$. This is the
result mentioned at the end of Section~\ref{sec:GLE}.  

Let us start now from the memory kernel which includes a subleading correction at long times:
\begin{eqnarray}
\gamma(t) \simeq \frac{k_B T}{a^2} t^{-2/z} 
\left[1 - \nu t^{-\frac{\nu+1}{\nu z} 
+ \frac{2}{z}}\right]
\label{app:defgamma}
\end{eqnarray}
where the time is made dimensionless with the unit $\tau_0$.
Its Laplace transform is:
\begin{widetext}
\begin{equation}
\hat{\gamma}(s) = \frac{k_B T}{a^2} \left[ 
\Gamma\left(1-\frac{2}{z}\right) s^{2/z-1} - 
\nu 
\Gamma\left(1-\frac{\nu+1}{\nu z}\right) 
s^{\frac{\nu+1}{\nu z}-1}\right] 
=\frac{k_B T}{A a^2}  s^{2/z-1} \left[1-B s^{\frac{\nu+1}{\nu z}-2/z}
\right]
\label{eq:hatgamma}
\end{equation}
\end{widetext}
where we have introduced
\begin{equation}
B\equiv \nu 
\frac{\Gamma(1-\frac{\nu+1}{\nu z})}{\Gamma(1-2/z)}>0 
\end{equation}
{and}
\begin{equation}
A^{-1}\equiv  \Gamma(1-2/z)>0
\end{equation}

%
%
%
%
%
%
%
%
%
From (\ref{eq:hatmu}) and (\ref{eq:hatgamma}) we get
\begin{eqnarray}
\hat{\mu}(s)=\frac{a^2}{k_B T} A s^{1-2/z} 
\left[1-B s^{\frac{\nu+1}{\nu z}-2/z}  \right]^{-1}
\label{mus}
\end{eqnarray}

The inverse Laplace transform can be calculated using the Mittag-Leffler
function \cite{haub11}. However, in order to avoid possible convergence
issues we will only calculate $\mu(t)$ in the long time limit, which
corresponds to the small $s$ approximation of (\ref{mus}). In that limit
we get
\begin{eqnarray}
\hat{\mu}(s) = \frac{a^2}{k_B T} A \left[ s^{-\kappa} + B s^{-\epsilon-\kappa} + \ldots \right]
\label{musa}
\end{eqnarray}
where $\epsilon=2/z-(\nu+1)/(\nu z) < 0$ and $\kappa=2/z-1 < 0$.
The inverse Laplace transform of (\ref{musa}) is given by
\begin{eqnarray}
\mu(t) = \frac{a^2}{k_B T} A \left[ \frac{t^{\kappa-1} }{\Gamma(\kappa)} + B \frac{t^{\epsilon+\kappa-1}}{\Gamma(\epsilon+\kappa)}  \right]
\label{9}
\end{eqnarray}



\subsection{Mean squared displacement of the reaction coordinate}

In absence of forces Eq.~(\ref{GLE_2}) becomes
\begin{eqnarray}
\vec{v}(t) = \vec{\eta}(t)
\label{1}
\end{eqnarray}
where 
\begin{eqnarray}
\langle\eta_{\text i}(t)\eta_{\text j}(s)\rangle
=k_BT \mu(|t-s|) \delta_{\text{ij}}
\label{2}
\end{eqnarray}

From (\ref{1}) we get for the mean squared displacement (MSD)
$\Delta \vec{x}^{\, 2}(t) = \langle (\vec{x}(t)-\vec{x}(0))^2\rangle$
\begin{eqnarray}
\Delta \vec{x}^{\, 2} (t) = 3 \int_0^t \int_0^t \langle \eta(t_1) \eta(t_2)\rangle dt_1 dt_2
\end{eqnarray}
Therefore using (\ref{2}) and (\ref{9}) we get
\begin{widetext}
\begin{eqnarray}
\Delta \vec{x}^{\, 2} (t) = 3 a^2 A \int_0^t \int_0^t  
\left[ \frac{|t_1-t_2|^{\kappa-1}}{\Gamma(\kappa)}  + B
\frac{|t_1-t_2|^{\epsilon+\kappa-1}}{\Gamma(\epsilon+\kappa)}  \right] dt_1 dt_2
\label{11}
\end{eqnarray}

\end{widetext}

Integrals of the type 
\begin{equation}
I=\int_0^t \int_0^t |t_1-t_2|^{\sigma} dt_1dt_2
\end{equation}
are easily performed. First, from the symmetry between $t_1$ and $t_2$ we get
\begin{equation}
I = 2 \int_0^t (\int_0^{t_1} (t_1-t_2)^\sigma dt_2) dt_1
\end{equation}
and then switching to $y=t_2/t_1$ 
\begin{equation}
I=2 \int_0^t t_1^{\sigma+1} dt_1 \int_0^1 (1-y)^\sigma dy=
\frac{2t^{2+\sigma}}{(2+\sigma)(1+\sigma)}
\label{13}
\end{equation}
provided $\sigma > -1$. Otherwise we get a divergence at the
origin. However, physically we can always introduce a small cutoff and
take the initial time to be some small time $t_\epsilon$. This will add
a constant to the result (\ref{13}).

Inserting (\ref{13}) into (\ref{11}) we get 
\begin{widetext}
\begin{eqnarray*}
\Delta \vec{x}^{\, 2} (t)= 6 a^2 A \left[\frac{t^{\kappa +1}}{\kappa(\kappa+1)\Gamma(\kappa)}   +
B \frac{t^{1+\kappa +  \epsilon}}{\Gamma(\epsilon +\kappa)(\kappa+1+\epsilon )(\kappa+\epsilon )} 
+ \ldots \right]
\end{eqnarray*}
Using the relation $z\Gamma(z)=\Gamma(z+1)$, 
\begin{eqnarray}
\Delta \vec{x}^{\, 2} (t) = 6a^2 A\left[\frac{t^{1+\kappa}}{\Gamma(\kappa+2)} + 
B \frac{t^{\epsilon+\kappa+1}}{\Gamma(\kappa+\epsilon+2)} + \ldots\right]
\label{15}
\end{eqnarray}
\end{widetext}

We next go from motion in physical space to motion in monomer space
throught the relation $\Delta \vec{x}^{\, 2} (t) \sim [\Delta
n^2(t)]^\nu$. Inserting in (\ref{15}) gives our final result from which
one can read off the leading correction to the asymptotic scaling
\begin{equation}
\Delta n^2(t) \sim t^{(1+\kappa)/\nu}
\left[1 + \frac{B\Gamma(\kappa+2)}{\nu \Gamma(\kappa + \epsilon+2)} t^\epsilon + \ldots\right]
\end{equation}
Inserting $\nu=.588$ then gives
\begin{equation}
\langle \Delta n^2(t) \rangle \sim t^{.92}\left[1 + C\,t^{-0.19} + \ldots\right]\label{final}
\end{equation}
where the prefactor $C$ in front of the correction is positive, whose value is calculated as $C\approx1.695 B\approx1.726$
 for the present case ($z=1+2\nu$ with $\nu=0.588$).

\section{Numerical simulations}
\label{sec:sim_num}

The numerical model used in this article was also used in studies of renaturation dynamics \cite{ferr10} and zipping dynamics \cite{ferr11,walt12}.
 The system is composed by two polymers defined on a face-centered-cubic lattice.
 The monomers on both strands are labeled with an index $i =$0, 1,\dots,$N$ where $0$ is the label of the free ends and $N$ 
the label of the opposite ends, see Fig.1(b). The two strands are self- and mutually avoiding, with the exception of
 monomers with the same index $i$, which are referred to as complementary monomers.
 Two complementary monomers can thus overlap on the same lattice site and bind to each other.
 In the starting configuration of Fig.2, the two strands are bound for $N/2\le i\le N$ and unbound for $i<N/2$. 
In Fig.3, we checked the scaling of the (un)zipping time with two other initial conditions with strands bound for $N/4\le i\le N$ and $3N/4\le i\le N$.
 This initial configuration is relaxed to equilibrium by means of pivot moves \cite{madr88} consisting in rotating a whole branch of polymer at once. 
 These pivot moves leave the number of bonds unchanged and are applied to both double and single stranded parts of the polymer. 
Given the length of polymer considered ($N\le768$), this equilibration is negligible compared to the sampling time needed to 
probe the dynamics of the reaction coordinate with a local algorithm.
 The simulation is started after equilibration, where the polymers undergo Rouse dynamics which consists
 of local corner-flip or end-flip moves that do not violate self- and mutual avoidance.
 The overlap between complementary monomers, which thus form a bound pair, is always accepted as a move.
 The opposite move of unbinding two bound complementary monomers is accepted with probability $\omega^{-1}=\exp(-\epsilon/k_B T)$, 
in agreement with detailed balance condition. Here the energy units are expressed in unit of the thermal energy $k_BT$, 
where $k_B$ is the Boltzmann constant and $T$ the temperature. An elementary move consists in selecting a random monomer
 on one of the two strands. A unit of time is defined as $N$ such random attempts of corner flip, i.e., a sweep of the polymer.
 If the selected monomer is unbound a local flip move is attempted. If the selected monomer is a bound monomer there are two
 possibilities. Either a local flip of the chosen monomer is attempted, and if accepted, this move results in the bond breakage;
 or a flip move of both bound monomers is generated, which does not break the bond between them. In the model discussed here we
 do not allow any bubble formation neither for zipping nor unzipping, by imposing the constraint that monomer $i-1$ can bind to its complement
 only if monomer $i$ is already bound. Analogously monomer $i+1$ can unbind only if monomers $i$ are already unbound.
 This is the model Y which was referred to in Ref.\cite{ferr11}.

\bibliography{refs}

\end{document}